\documentstyle [12pt]{article}
\begin{document}
\newcommand{\vm}{\vspace{0.2cm}}
\newcommand{\vl}{\vspace{0.4cm}}

\title{p-Adic  description of Higgs mechanism V: New Physics }
\author{Matti Pitk\"anen\\
Torkkelinkatu 21 B 39, 00530, Helsinki, FINLAND}
\date{6.10. 1994}
\maketitle
\newpage

\begin{center}
Abstract
\end{center}

\vm

This is the fifth  paper in the series devoted to the calculation particle
and hadron masses in the p-adic field theory limit of TGD. In this paper
the
 possibility of two new branches of physics suggested by TGD, namely
$M_{89}$
 hadron
 physics and $M_{127}$ leptohadron physics,
 is considered.  According to  TGD  leptons and U type quarks have
colored  excitations. The anomalous production of $e^+e^-$ pairs in heavy
ion collisions  indeed suggests the  existence of
 light leptomeson decaying to $e+e^-$ pairs.  There are however grave
objections against the existence of light exotics.
 First,  asymptotic freedom in the standard sense is  lost unless the
exotic colored bosons with spin one  save the situation.
 Secondly,   $Z^0$ decay widths seem to exclude  light exotic   fermions.
The  solution of the problem is based on p-adic unitarity and probability
concepts. In general the real counterpart for the sum of p-adic
probabilities differs  from the sum for the  real counterparts of
individual probabilities.
 The interpretation is that the  sum of p-adic decay  probabilities
corresponds to the situation,  where individual final states are not
monitored separately but a common signature for final states is used
whereas the sum of real  probabilities describes the situation,  where
each final state is  monitored separately.
 An elementary consequence of p-adic unitarity is that  the  p-adic
transition rate  for $X \rightarrow anything$   vanishes for   any initial
state
 (as it must since there is no signature for 'anything ') ,  when
indivudial final states are not monitored.  The total  decay rate of $Z^0$
to unmonitored exotic leptons (p-adic sum of  probabilities!) is
 very sensitive to the value of $sin^2(\theta_W(eff))$  and  indeed
vanishes for  $sin^2(\theta_W(eff))=0.23114$!  The decays   $ \mu
\rightarrow \pi_L$ and $n\rightarrow p+\pi_L$ afford a
 unique possibility to detect charged  leptopion and one obtains upper
bound  for  $W-\pi_L$ coupling. TGD  predicts no Higgs particle and
$M_{89}$ hadron physics  obtained by scaling the old hadron physics mass
scale by factor $512$  seems to be the TGD:eish counterpart of Higgs
necessary  to
 make theory unitary.
  The anomalies  reported in the production and decay of top quark
candidate might result from the production of  $u$ and  $d$  quarks
of $M_{89}$ hadron physics having masses very nearly equal to the masses of
top candidate. Therefore the   signatures of the
$M_{89}$ Physics are considered.  The concept of topological  evaporation
suggests the identifaction of Pomeron as sea in TGD picture.

\newpage

\tableofcontents

\newpage

\section{Introduction}

p-Adic TGD makes possible surprisingly detailed understanding of
elementary  particle and hadron masses. Although the hadronic mass
formula   contains several parameters, which cannot be predicted at this
stage one can deduce the values of these integer parameters by number
theoretic constraints plus few empirical
 inputs. TGD differs from standard model in some respects. \\ a) TGD
predicts also some exotic fermions and bosons.
Color  decuplets ( $10,\bar{10}$) of  charged and neutral leptons and
 $27$-plet
of  colored neutrinos as well as color excitations of $U$ type
quarks created by color decuplets  are predicted. Also exotic colored
 bosons
are predicted in representations $8,10,\bar{10}$ and $27$.  This suggests
that   asymptotic freedom in standard sense is lost unless spin one
colored bosons save the situation.  The decay width of $Z^0$  boson seems
 also
to exclude new light fermions.  The observation of $e^+e^-$ pairs in heavy
ion
 collisions, which seem to originate from unknown pseudoscalar meson of
mass of order  $MeV$ suggest  together with esthetic arguments suggest
that leptohadrons exist. This suggests the possibility  that p-adic
unitarity and probability concepts migt indeed allow leptohadron physics.
\\ b) In previous
 paper it was found that hadron masses can be understood within one per
cent errors and even
 isospin splittings can be understood. The only exception was top quark,
whose mass  for $k(top)=89$ ($k(top)=97$) was predicted to
 be about five times larger  (3 times smaller) than the mass of the
 observed top candidate \cite{Abe}. The properties of CKM matrix seem to
force the identification of the top candidate as actual top and
 small mixing of
$k=97$ and $k=89$  condensate levels gives required mass for top quark.
The mass of top candidate is however quite close to the masses of $u$ and $d$
quarks of $M_{89}$ hadron physics and the failure to distinguish between
the actual top and $u_{89}$ and $d_{89}$ might explain  the reported
anomalies  in production and decays of the top candidate.   \\ c) The
existence of  $M_{89}$  hadron physics is suggested by the unitarity
requirement also.  There is no room for Higgs in TGD and $M_{89}$ hadrons
would be the counterpart of Higgs needed to guarantee unitarity.

\vm

In the following the possibility  of leptohadron physics is considered. \\
a) It is shown that asymptotic freedom is not lost: this is due to
the existence of $J=1$ exotic colored bosons whose contribution dominates in
$\beta$ function.
  The existence of  QCD for each Mersenne
prime is proposed and color coupling strength $\alpha_s (M_n)$
 becomes large
(of order $p-1+O(p)$) for $\Lambda (M_n)$.
 Below $(\Lambda (M_n)$ modulo mathematics
saves the situation since p-adic color couplings  strength as well as beta
function are of order $O(p)$ and have extremely small real counterparts,
which in
 practice means the end of that particular QCD.
 \\
b)  The problem of $Z^0$ decay widths forces the detailed study of p-adic
 unitarity and probability concepts and leads to  profoundly new ideas
 concerning physical  measurement. The point is that the real counterpart
for the sum of  p-adic probabilities is not identical with the sum for the
real counterparts of individual  probabilities. The interpretation is as
follows: the p-adic sum of probabilities applies to
 the measurement,   where only a common signature for final states is used
whereas the sum for  real  counterparts of the individual  probabilities
refers to the situation,  where each final state is monitored
 separately using a spesific signature.  The idea that measurement
situation affects the outcome in this manner is new although physically
there is no mysterious in it since the experimental  arrangements are
competely different. \\ c) A particular consequence  of the  p-adic
unitarity is that the p-adic
 probability  for $X\rightarrow anything$
 for a given initial state  $X$ vanishes.  The explanation is simple: there
 is no experimental signature for 'anything' and therefore even the test
of this prediction is
 impossible.
   \\ d) The p-adiction of decay amplitudes squared for gauge bosons shows
that it  total
 p-adic probability for $Z^0$ to decay into exotic leptons in final state
can be  made very small with the finetuning of the  Weinberg angle: also
now modulo arithmetics
 is at work. \\ e)  The decays $\mu \rightarrow \nu_{\mu}+\pi_L$ and
$n\rightarrow p+\pi_L$  provide unique tests for leptopion hypothesis and
expressions for decay rates and bounds for  appropriate couplings using CVC
and PCAC hypothesis.

\vm

The second part of the paper is devoted to the   newly interpreted
 top quark candidate and the possible signatures of $M_{89}$ hadron
physics.  As found in previous paper CKM matrix seems to favour the
identification of the observed top candidate as actual top and one can
understand top mass  if  small condensate level mixing between
$k=97$ and $k=89$ levels takes place. The mass of top candidate is nearly the
same as the masses of u and d quarks of $M_{89}$ Physics. Therefore one can
consider the possibility that the production of $M_{89}$ hadrons could explain
the reported anomalies in top production rate.  Topological
evaporation concept suggests that  the newly born  concept  of Pomeron of
Regge theory  can be identified  as the sea of perturbative QCD.
 The possibility to  identity these  basic phenomenological concepts
shows the  unifying power of TGD
 framework.

\section{Exotic  bound states  of leptons}

TGD predicts colored  excitations for leptons and in \cite{Lepto,Heavy}
 an explanation for the anomalous production of $e^+e^-$ pairs in the
collisions of heavy nuclei  based on the concept of leptopion was
proposed.  Leptopion,  and more generally, leptohadrons were assumed to be
bound states of color octet
 excitations of  leptons and it was found that generalization of PCAC
hypothesis gives even  quantitatively satisfactory description for the
anomalous $e^+e^-$ production.  It is useful to reconsider the hypothesis
in light of recent experimental
 facts.

\vm

\noindent a) Empirical constraints do not force the constituents of the
resonance  decaying to  $e^+e^-$ pair to be colored: what is however
needed is some confinement  mechanism allowing excitations with higher
mass. If constituents are colored they need not be
 necessarily  color octets as originally believed to be.  If one
identifies the resonance as color bound state there are two natural
possibilities: the identification as either leptopion or as
lepto-$\eta$.  \\ b) A strong empirical constraint on the number of light
massless fermions
 comes from the decay widths of intermediate gauge bosons. The decay width
is simply the  sum of contributions coming from various light fermions and
$Z^0$ decay width doesn't allow new light fermions. It is difficult to
imagine why massless colored leptons should
 not be regarded as mass elementary fermions and it seems that one must
assume that  colored leptons and quarks  are condensed on level with $p\ge
M_{89}$ and one would  lose the elegant explanation of anomalous $e^+e^-$
pairs.   \\ c) Asymptotic freedom  is a sacred hypothesis of  practically
all unified model building and one must find whether the exotic colored
$J=1$ bosons can  preserve asymptotic freedom threatened by the
existence of exotic fermions.
  \\ d)
One can consider  also a  rather science fictive possibility that p-adic
probability might pose only a weaker condition on the numbers of  various
light fermions via $Z^0$ decay widths: condition would which the numbers
only modulo certain integer so that color excited quarks and leptons could
exist peacefully.  The consideration of this admittedly crazy idea turns
out to be very fruitful since it leads to understanding of the
relationship between p-adic and real unitarity and proability concepts and
gives some foretaste of  how deep role  number theory plays  in the
construction of  S-matrix.   Even more: explicit calculations show that
p-adic loophole  exists!

\subsection{ Relationship between p-adic and real probabilities}

p-Adic quantum field theory gives rise to transition probabilities
 $P_{ij}$, which are p-adic numbers.   The first problem is to associate
real conserved probabilities to p-adic probabilities.  The
identification   is based on  a simple renormalization for the real
counterparts of the p-adic probabilities  $(P_{ij})_R$ obtained
 by canonical identification.

\begin{eqnarray}
P_{ij}&=& \sum_{k\geq 0} P_{ij}^k p^k\nonumber\\
P_{ij} &\rightarrow & \sum_{k\geq 0} P_{ij}^kp^{-k}\equiv (P_{ij})_R
\nonumber\\
(P_{ij})_R&\rightarrow & \frac{(P_{ij})_R}{\sum_j (P_{ij})_R}
\equiv P_{ij}^R
\nonumber\\
\
\end{eqnarray}

\noindent  The  procedure converges rapidly  in powers of $p$ and is
highly reminiscent of renormalization of quantum field  theories
\cite{padTGD}.   The renormalization  procedure  automatically divides
away one four- momentum delta function from the square of S-matrix element
containing square of delta function with no well defined  mathematical
meaning. Usually one gets rid of the deltafunction interpreting it as the
inverse of the four-dimensional measurement volume so that transition rate
instead of transition probability is obtained. Of course, also now same
procedure should work either as a discrete or continuous version.

\vm

 The crucial question is what is the physical difference between the real
counterpart for sum of p-adic  transition probabilities and for the sum
of the real counterparts of these probabilities, which are in general
different

\begin{eqnarray}
(\sum_jP_{ij})_R &\neq& \sum_j (P_{ij})_R
\end{eqnarray}

\noindent  There is also a problem of renormalization. The suggestion is
that
 p-adic sum of transition probabilities corresponds to the  experimental
situation, when one does not  monitor individual transitions  but using
some common experimental signature   only looks  whether the transition
leads to  this set of finals states or not. When one looks each
transition  separately or effectively performs different experiment by
considering only one transition  channel in each experiment one
 must use the sum of real probabilities.  More precisely the choice of
experimental  signatures divides the set $U$ of the  final states to
disjoint union $U=\cup_i U_i$ and one must define the real  counterparts
for transition probabilities $P_{iU_k}$ as

\begin{eqnarray}
P_{iU_k}&=& \sum_{j\in U_k} P_{ij}\nonumber\\
P_{iU_k} &\rightarrow &  (P_{iU_k})_R\nonumber\\
(P_{iU_k})_R&\rightarrow & \frac{(P_{iU_k})_R}{\sum_l (P_{iU_l})_R}
\equiv P_{iU_k}^R\nonumber\\
\
\end{eqnarray}

\noindent   The assumption  means deep difference with respect to the
 ordinary probability theory.    Physically there is nothing mysterious in
the difference since the experimental situations are quite
 different in two cases.   The procedure is even familiar for physicsts!
Assume that  the labels $j$
 correspond to  momenta.
 The division of  momentum space to cells of given size so that individual
momenta inside  cells are not monitored separately  means that momentum
resolution is finite. Therefore one must  perform p-adic sum over cells
and define real probabilities in the proposed manner.  p-Adic effects
  resulting from  the difference between p-adic and real summation are the
counterpart  to renormalization effects in QFT.  It should be added that
similar resolution can be defined also for initial states by decomposing
them into a union disjoint subsets.

\vm

 p-Adic probability conservation implies that the lowest order terms for
 p-adic probabilities satisfy the condition $\sum_jP^0_{ij}=1+ O(p)$.  The
general
 solution to the condition is $P^0_{ij}=n_{ij}$. If the number of the
final states is
 much smaller than $p$ this alternative implies that  real transition
rates are enormous:
 typically of order $p$! Therefore it seems that one must   assume

\begin{eqnarray}
P^{0}_{ij}&=&\delta (i,j)
\end{eqnarray}

\noindent
  As a consequence  the probability for anything  to happen (no monitoring
of different  events)  is given by

\begin{eqnarray}
\sum_j(P_{ij}-\delta (i,j))&=&0
\end{eqnarray}

\noindent and  vanishes identically!
   This is not so peculiar as it looks  first since there must be some
signature for anything to happen in order that it can be
 measured and signature always distinguishes between two different events
at least: it  is difficult to imagine what the statement  'anything did
not happen' might mean!  Of course, in real context this philosophy would
imply the triviality of
 S-matrix.

\vm

The unitarity condition implies   that the the moduli squared of the
matrix $T$ in $S=1+iT$ are of order
 $O(p^{1/2})$  if one assumes four-dimensional p-adic extension allowing
square root  for p-adically real numbers and  one can write

\begin{eqnarray}
S&=&1+i\sqrt{p}T\nonumber\\
i(T-T^{\dagger})+\sqrt{p}TT^{\dagger}&=&0
\end{eqnarray}

\noindent It must be emphasized that this expression is completely
analogous
 to the ordinary one since $i\sqrt{p}$ is one of the units of the
four-dimensional  algebraic extension.  Unitarity condition  in turn
implies recursive solution of unitary condition in powers of $p$:

\begin{eqnarray}
T&=&\sum_{n\ge 0}T_np^{n/2}\nonumber\\
T_n-T_n^{\dagger}&=&\frac{1}{i}\sum_{k=0,..,n-1} T_{n-1-k}T_{k}^{\dagger}
\end{eqnarray}

\noindent  If  algebraic extension is not allowed then the expansion is in
powers of $p$ instead of $\sqrt{p}$. Note that the real counterpart of the
series convergenges extremely rapidly.

\vm

Before one is able to p-adicize decay rates of gauge bosons one must solve
 some problems related to the transition from p-adic to real realm.  In
order to  obtain transition rates one must integrate p-adically over the
final state momenta.
 The ordinary real integration measure

\begin{eqnarray}
dV_f&=& (2\pi)^4\prod_{i=1}^{N_f}\frac{ d^3p_i}{2E_i (2\pi)^3}
\end{eqnarray}

\noindent  is problematic due to the appearence of   powers of $\pi$  since
 it is not at all obvious what $\pi$ means p-adically.  The solution of
the problem is
 the necessary appearence of p-adic infrared cutoff, which  means that
summation is performed instead of integration and everything is
rational.     The decay rates contain also the  factor $1/2M(B)$ and this
need not be rational nor even p-adically real number. Same applies to the
flux factor appearing in cross sections.

\vm

 The practical  solution of these problems is based on physical argument.
One  has could reasons to
 argue that in good approximation final state momenta belong to state
labels,
 for which monitoring is performed for each momentum value separately so
that the integration over  the final state momenta must be  performed on
the real side so that the problems  related to powers of $\pi$  disapper
totally. In this case it is $\vert M \vert^2$, which is the correct p-adic
quantity to study and for a given momentum configuration one can still
perform p-adic sum over the   degrees of freedom not monitored.  Assuming
that the squares $g_i^2$  of coupling  constants and other analogous
parameters  are rational numbers and also that  four-momenta are rational
(by the necessary infrared cutoff)  $\vert M \vert^2$ is indeed rational
number.

\vm

The proposed interpretation for p-adic probabilities suggests a  new
formulation of  renormalization theory. It is convenient to use spacetime
picture, spacetime points being defined as p-adic space dual to the
momentum space.  Finite resolution means division of
 the p-adic spacetime  into cells of given size so that no monitoring of
individual points  inside cell is performed: this implies
 effective momentum cutoff. A nice feature is that if cell size is defined
using p-adic  norm space time decomposes into disjoint regions in p-adic
case (ultrametricity
 \cite{padTGD}).  The
 theories associated with  different cell sizes correspond  to different
physical situations  and renormalization is therefore not a mere
calculational  tool.
 Transition rates are obtained by summing the p-adic transition
probabilities p-adically over each  cell associated with final state
spacetime points,
 averaging over
 initial  state space time points  over each cell,  mapping the result to
reals and performing
 the needed renormalization for the real probabilities. Momentum space
description is obtained by taking Fourier transform for each incoming and
outgoing particle.
 Infinitesimal renormalization operation  corresponds to the  change in
transition
 probabilities induced by a change in the resolution scale so that
generalized beta functions can be
 defined  in terms of the derivatives with respect to  the length scale
for the real
 counterparts of  the integrals over momentum space cell volume.    The
real valued running  couplings constants involving logarithms
$ln(Q^2/Q^2_0)$   result basically from the change of the physical
situation,  when the length scale $L=1/Q$, below  which no monitoring is
carried out changes.   p-Adically the logarithms do not even make sense
for all momenta.

\vm

p-Adic probability concept provides  an elegant solution to the infrared
divergence problems of  the gauge theories. The  measurement of
scattering rates of  say two charged particles in forward directions,
where real transition rates diverge  and lead to infite  cross section in
real QED,  is limited by angular resolution. This  means that one cannot
monitor individual events in certain small cone around forward  direction
and total rate  for this small cone is obtained by summing p-adic
probabilities  for the cone in question and mapping the sum to the
reals.   The total  p-adic rate  is however always of form  $P(if)=
Xp+...$,  where  $X$  cannot increase above $p$ so
 that total rate  remains finite.

\subsection{ Do colored excitations  imply  the loss of asymptotic freedom?}

The most spectacular consequence of the existence of colored excitations
 would be  the loss of asymptotic freedom in the conventional sense of the
 word
 for  both leptohadrons and ordinary hadrons.   In the real theory  the
beta
 function corresponds
 to the sum over fermion and gluon loops for gluon propagator

\begin{eqnarray}
\alpha_s(Q^2)&\simeq& \frac{\alpha(Q_0^2)}
{(1+ \alpha_s(Q_0^2)\frac{b_0}{4\pi}ln(Q^2/Q_0^2)  )}\nonumber\\
b_0 &=& \frac{11}{6}l(J=1)- \frac{2}{3}l(J=1/2)-\frac{1}{12}l(J=0)
\nonumber\\
l(J)&=& \sum_{ light \ ( D,J)} l(D,J)
\end{eqnarray}

\noindent Where the total index $l(J)$ associated with spin $J$ particles
is sum over the indices of color representations $D$ with spin $J$.
 Fermions are assumed
to be Dirac fermions.
Summation is over  the colored elementary particles, which are light in
 mass scale $Q_0^2$.

\vm

  The list for the values of the indices
$l(D)$  for the  representations in leptonic ($10,\bar{10},2\times 27$)
 and quark sector $3\otimes (10+\bar{10})= 15+15'+6+24$  is
given  in table below.  The second table lists exotic colored bosons for
$T=1$ and $T=1/2$ and the remaining tables give the contribution of single
exotic generation to $b_0$ and asymptotic value of $b_0$ for $T=1$ and
 $T=1/2$.

\vl

\begin{tabular}{||l|l|l|l|l|l|l|l||}\hline \hline
lepton \ N& $n(N)$&$l(N)$&$C(N)/C(3)$& quark  \ N
&$n(N)$&$l(N)$& $C(N)/C(3)$\\  \hline\hline
10& $1$&15&$\frac{9}{2}$ &15 &$1$&20 &4 \\ \hline\hline
$\bar{10}$&$1$&15&$\frac{9}{2}$  &15' &1&35&7\\ \hline\hline
$27$&2& 54&  6& $6$&$1$&5&$\frac{15}{2}$\\ \hline\hline
& && &24 & $1$&50&$\frac{25}{4}$\\ \hline\hline
\end{tabular}

\vl

Table 2.1. \label{Colorcasimir} The degeneracies $n(N)$ of color
 representations and so
called
index $l(N)$ of the representation for the exotic leptons and  quarks.
 The  table contains also the values of Casimir operators needed
later.

\vl

\begin{tabular}{||c|c|c|c|c|c||}
 \hline\hline
spin  &charge operator & $D$&$n(D)$ &$ M^2 (T=1)$&$M^2(T=1/2)$ \\
\cline{1-6}\hline
0& $I^{\pm}Q_K$ &8&1 &$\frac{2}{p}$&0\\ \hline
1& $I^{\pm}$ &8 &1&Planck mass&$\frac{1}{2p}$\\ \hline
1& $I^3_{L/R}Q_K$ &8 &1&Planck mass&$\frac{1}{2p}$\\ \hline
1&$I^{\pm}Q_K$&8 &1&Planck mass &$(\frac{3p}{10})_R\frac{1}{p}$
\\ \hline\hline
0& $I^{\pm}Q_K$ &$10,\bar{10}$ &1&$\frac{3}{p}$&0\\ \hline
0& $1$ &$10,\bar{10}$ &1&0&$0$\\ \hline
1&$I^{\pm}$&$10,\bar{10}$ &1&$\frac{3}{p}$&0\\ \hline
1&$I^3_{R/L}Q_K$&$10,\bar{10}$ &1&$\frac{3}{p}$&0\\ \hline\hline
0&$I^{\pm}$&27 &1&Planck mass &$\frac{1}{2p}$\\ \hline
0&$I^3_{R/L}Q_K$&27 &1&Planck mass &$\frac{1}{2p}$\\ \hline
1&$I^3_{R/L}$&27 &1&Planck mass &$\frac{1}{2p}$\\ \hline
1&$Q_K$&27 &1&Planck mass &$\frac{1}{2p}$\\ \hline\hline
0&$I^3_{R/L}$&27 &2&0 &0\\ \hline
0&$Q_K$&27 &2&0 &0\\ \hline
1&$1$&27 &2&0 &0\\ \hline\hline
\end{tabular}

\vl

Table 2.2. \label{Colbosonmasses} Charge operators, degeneracies and masses
 of colored
 exotic
bosons for p-adic temperature $T=1$ and $T=1/2$.  The last two massless
 bosons are doubly
 degenerate due to occurrence of two $27$-plets with conformal weight
$n=2$.

\vl

\begin{tabular}{||l|l|l|l||}\hline \hline
T & $\Delta b_0(B)$ &$\Delta b_0 (q)$& $\Delta b_0(L)$\\
 \hline
1& $ 230-1/2$ &-112&  $-\frac{220}{3}$  \\ \hline
$\frac{1}{2}$& $ 639-1/2$ & -112 & $-\frac{220}{3}$
\\ \hline\hline
\end{tabular}

\vl

\noindent Table 2.3. \label{Beta} Contributions of single exotic fermion and
boson family to the coefficient $b_0$ for p-adic temperature $T=1$ and
$T=1/2$.

\vl

\begin{tabular}{||l|l|l|l||}\hline \hline
T & n(boson  family) &$b_0 (h)$& $b_0(L)$\\ \hline
1& 1&$11-\frac{2n_f}{3}+ 230-\frac{1}{2}- 220$ & $11+ 230-1/2- 336$ \\ \hline
1&3&$11-\frac{2n_f}{3}+ 689-\frac{1}{2}- 220$ & $11+ 689-1/2- 336 $ \\
 \hline\hline
$\frac{1}{2}$&1& $11-\frac{2n_f}{3}+ 639-1/2- 220$ & $11+ 639-1/2- 336$
\\ \hline
$\frac{1}{2}$&3& $11-\frac{2n_f}{3}+ 1916-1/2- 220$ & $11+ 1916-1/2- 336$
 \\ \hline\hline
\end{tabular}

\vl

\noindent Table 2.4. \label{Asympt} Asymptotic values of $b_0$ for
hadronic and leptohadronic gluons for $T=1$ and $T=1/2$ and for
single and 3 exotic  boson families respectively.
 The standard contribution to $b_0$ is written separately.

\vm

Some general  conclusions can be made by inspecting these tables.\\
a) It seems natural to assume  $\cite{padTGD}$ that each different
 QCD (one for each
 Mersenne prime)
has its own gluons and
colored bosons and that confinement makes the direct communication between
different  QCD:s impossible (gluons of $M_{8107}$ QCD do not couple to
 quarks of $M_{89}$ QCD).  The assumption implies that colored lepton
 pairs are not produced in hadronic
 reactions via gluon emission.
\\
b) In TGD one expects family replication phenomenon for bosons, too.
All $g>0$ bosons are massive and the modular contribution to mass squared
is
same as for fermions and dominates.  Family replication implies asymptotic
freedom even for $T=1$. $T=1/2$ alternative is probably excluded
since it implies too rapid coupling constant evolution.
 For $T=1$ asymptotic freedom
 achieved in leptonic sector with  2 boson generations whereas one boson
family
is enough in hadronic sector.  Coupling constant evolution is sensitive to
the masses of exotics and therefore to the topological mixing  of exotic
fermions and bosons.
\\
c)  The massless $(J=1,g=0)$  $27$-plet gives
the dominating $J=1$ contribution to beta function and implies extremely
 rapid
coupling constant evolution. In case of hadrons this possibility is
definitely excluded.
 Catastrophe is avoided  if
$g=0$ multiplet suffers topological mixing with $g>0$ 27-plets and becomes
massive.
\\
 d)  In case of hadrons there seems to be no place for light exotics.
The first possible primary condensation level for exotic
bosons and quarks seems to be
$M_{89}$: otherwise the  exotic bosons and fermions should have been seen  in
$e^+e^-$ annihilation to hadrons: also hadrons having exotic quarks
as constituents would have been observed. Color coupling evolution becomes
very rapid after first exotic boson generation unless all exotic fermions
and first generation bosons have nearly identical masses. Note that the
masses
 of
first generation exotic $U$  quarks  suffered primary condensation  on
$M_{89}$ level have mass essentially identical to the mass of the observed
top candidate and could also contribute to the signal besides $u$ and $d$
 quarks
of the $M_{89}$ hadron physics.

\vm

The previous considerations are based on experience
with  ordinary real QFT.   The problem is to understand what is the p-adic
counterpart of the real
 coupling constant evolution.\\ a)   p-Adic logarithms  $ln(Q^2/Q_0^2)$
appearing in the expressions of running coupling constants  exist only
provided  the condition

\begin{eqnarray}
\frac{Q^2}{Q_0^2}&=&1+ O(p)
\end{eqnarray}

\noindent holds true. This suggest that in p-adic theory one has distinct
 coupling constant evolution in each cell

\begin{eqnarray}
Q^2&=&  Q^2(k,n)+O(p) \nonumber\\
Q^2(k,n)&=& np^k\nonumber\\
g^2(Q^2)&=& g(Q^2(k,n) f(ln(\frac{Q^2}{(Q^2(k,n)})))\nonumber\\
\end{eqnarray}

\noindent  The result implies  that there exist in principle two distinct
 coupling constant evolutions. One is  related to the changes of the
initial value
 $g^2(Q^2(k,n))$  as function of cutoff scale $Q^2((k,n)$  and one
evolution inside each  'cell' $(k,n)$. \\ b)  p-Adic conformal field
theory  describes critical system: this suggests  that
 beta functions should vanish identically and  coupling constant evolution
inside each cell $(k,n)$ ought to be trivial. This in turn would mean
that ordinary
 coupling constant evolution is replaced with stepwise jumps of the
coupling strenght  $g^2(k,n)$ and it is this evolution, which has  as real
counterpart  the ordinary logarithmic coupling constant evolution.    The
unpleasant consequence is that   that the argument leading to $Q^2(k,n)$
is lost. Cell structure  might well follow from the fact that p-adically
analytic functions of $Q^2$  appearing  in S-matrix elements converge only
inside cells $(k,n)$.
 \\ c)  The divergence of the coupling constant strength is not possible
in p-adic  theory. The largest possible value is achieved,  when  one has

\begin{eqnarray}
g^2_{max} &=& (p-1)(1+ p+p^2+...) =-1=p-1 +O(p)
\end{eqnarray}

\noindent What is remarkable is  that if evolution continues the next step
corresponds to  $g^2\rightarrow  g^2_{max}+1= O(p)$ so that the real
counterpart of the  coupling strength becomes extremely small, of the
order of gravitational coupling strength in  dimensionless units!  Also
beta function is of order $O(p)$ after this jump. This means the end  of
the QCD in question.

\vm

This general  picture fits  nicely into the concept of p-adic length scale
 hierarchy.
 Various QCD:s in the hierarchy of QCD:s correspond to Mersenne primes
$M_n$ and color coupling strength  becomes extremely small
below  $\Lambda (M_n)$.  One cannot exclude the possibility
that exotic bosons are much more massive than exotic fermions
and  «asymptotic freedom'  with extremely small color  coupling
strength sets on
 outside the finite range  $[\Lambda (M_n),lower), \Lambda (M_n,upper)]$.

\vm

 The fact that bosons determine coupling constant evolution near
 $\Lambda (M_n)$ plus fractality considerations suggest that
the scales $\Lambda (M_n)$ are universal and of form

 \begin{eqnarray}
\Lambda (M_n)&=& \frac{k}{\sqrt{M_n}}
\end{eqnarray}

\noindent   For leptonhadrons the hypothesis implies
 $\Lambda (M_{127})= 2^{-10}\Lambda_{QCD} \simeq 0.3 \ MeV$ for
$\Lambda_{QCD}\simeq 310 \ MeV$.   With same assumption one has
  $\Lambda (M_{89}) =
\frac{M_{89}}{M_{107}}\Lambda (QCD)  \simeq 158.7 \ GeV  $,
 which means that the observed top quark candidate
 could  belong to $M_{89}$ Physics.

\subsection{p-Adic loophole}

Intermediate gauge bosons exclude light exotic particles in standard model
and the predictions of the standard model are in excellent agreement with
experiments. One can ask whether p-adic probability concept might allow
room for new  light exotics.
 The point is that  in p-adic context the transition probabilities
$P(i\rightarrow j)$ satisfy the condition $\sum_{j}(P(i\rightarrow
j)-\delta (i,j))=0$ p-adically.    In case of decay rates this means that
system is stable if one only looks
 whether system decays but does not monitor specific decay channels.  When
one looks for
 specific decay channels the system is found to decay.  The  squares of
p-adicized decay amplitudes for intermediate gauge bosons are  indeed
  of order $O(p^2)$ so that the possibility that the total p-adic decay
rate to  unmonitored exotic channels  can be  very small and light exotic
leptons
 and quarks are therefore possible.

\vm

In order to see that  p-adic effects can solve the problem posed by gauge
boson decay widths,    one can  naively
 p-adicize the squares of the decay amplitudes of intermediate gauge
bosons:  by the simplicity of the process these quantities are  expected
to be quite universal in their dependence on various parameters.    The
explicit expressions for  the squares of the  decay amplitudes to the
standard channels are  in the approximation,
 when radiative corrections are not included  given by

\begin{eqnarray}
 \vert M (W\rightarrow L\bar{\nu}_L) \vert^2&=&
 \vert M (W\rightarrow U\bar{D}) \vert^2\equiv \vert M(W)
\vert^2\nonumber\\
&=&g_{ew}^2m(W)^2= g_{ew}^2 X(W)p^2 \equiv Y(W)p^2
\nonumber\\
 \vert M (Z\rightarrow \nu_e\bar{\nu}_e) \vert^2&\equiv&
\vert M(Z)\vert^2\nonumber\\
&=&
\frac{g_{Z}^2}{2}m(Z)^2= \frac{g_{ew}^2}{(1-P)^2} X(W)p^2\equiv Y(Z)p^2
  \nonumber\\
\vert M (Z\rightarrow e\bar{e} )\vert^2&=&  (1-4P+8P^2)\vert M(Z)\vert^2
 \nonumber\\
\vert M (Z\rightarrow U\bar{U} )\vert^2&=&  3(1-\frac{8P}{3})\vert
M(Z)\vert^2\nonumber\\
 \vert M (Z\rightarrow D\bar{D}) \vert^2&=&
3(1-\frac{4P}{3})\vert M(Z)\vert^2\nonumber\\
 P&=& sin^2(\theta_W)
\end{eqnarray}

\noindent  These quantities  are of order $O(p^2)$, which is due to the
fact  that gauge boson masses are of order $O(p^2)$. This implies that the
real decay  rates are of correct order of magnitude provided the
coefficient of $p^2$ is  rational
  number different from integer: integer part of this coefficient gives
totally neglible contribution to the
 decay rate.

\vm

  The occurrence of additional powers of $p$ is in accordance with the
 previous observation that $P^0_{ij}$ is nonvanishing for forward
scattering only.
 This means that that the sum of  p-adic decay  probabilities   is
vanishing.    Intermediate gauge bosons are p-adically stable against
decay to anything and unstable against  decay to monitored channels!
  This is possible if the denominators of the coefficients $Y(W)$ and
$Y(Z)$ are small integers so that modulo mathematics comes in game,  when
one  sums contributions over different non-monitored  channels p-adically.
This is
 indeed the case since the values of  $g_{ew}^2$ and $g_Z^2$ are of order
one $1/2$   and $X(W)$ and $X(Z)$ are apart from renormalization
corrections  given by $X(W)=1/2$ and  $X(Z)=1/5$.  This means that  the
sum over a fairly small number of unmonitored decay  channels gives
extremely small  real decay rate.

\vm

Consider next the mechanism at a  more detailed level. \\
 a) Weinberg angle and electromagnetic coupling $e^2$  appear in the decay
rate.  In calculating decay rates of gauge bosons one must use the
effective value of Weinberg angle taking into account various radiative
corrections.  From LEP precision experiments \cite{Fortschritte} one has
$P_{eff}\simeq 0.2324\pm 0.0005$. If
 $e^2 \simeq 0.091701$  rational number, which is a finite sum of finite
number of powers
 of two it behaves like real number in the canonical identification of
transition  amplitudes squared for Mersenne primes but not in general.  It
is quite conveivable
 that the real counterpart of  $e^2$ is much larger (or smaller) for
general p-adic prime  than for Mersenne primes and this perhaps  explains
why Mersenne primes are  in  favoured position dynamically. In a good
approximation
 one has $e^2= 47/512$ at low energies. In  intermediate boson mass scale
 the renormalized value of $\alpha_{em}(m(Z)^2) \simeq 128.87$, which in
good  approximation corresponds to $e^2= 50/512$,    implies sizable
correction to the decay rate.

\vm

\noindent b)   The numerical values of $Y(W)$ and $Y(Z)$ for LEP value
 $P_{eff}=0.2324$ and $e^2(m(Z)^2)$

\begin{eqnarray}
Y(W) &=& 0.2097\nonumber\\
Y(Z) &=& 0.1780
\end{eqnarray}

\noindent If  the   coefficients $Y(W)$ and $Y(Z)$ are in a sufficiently
good approximation finite sums of negative powers of two  the real
counterparts of decay rates  to single channel are near the values
predicted by standard model. It should be noticed  that the if the values
of $Y(B)$ would have been larger than one the p-adic scenario  would not
make sense since p-adic scenario predicts that the coefficients are not
larger than one. From the values of $Y(W)$ and $Y(Z)$ its clear that the
p-adic effects  become important,  when the number of individually
unmonitored  channels is larger  than $5$ so that p-adic modulo effects
are bound to affect crucially the total decay  rates of $Z$ and $W$.

\vm

\noindent   c) The naive  manner  to take  into account QED and QCD
renormalization corrections is to p-adicize  the corrections predicted by
real theory

\begin{eqnarray}
 \vert M (W\rightarrow L^{10}\bar{\nu}_L^{10}) \vert^2&=&
 10 X(10)   \vert M (W\rightarrow e\bar{\nu}_e) \vert^2\nonumber\\
 \vert M (Z\rightarrow  F^{N}\bar{F}^{N})\vert^2  &=& NN(Q(F),N,\alpha_s)
\vert M
(Z\rightarrow  F\bar{F})\vert^2\nonumber\\
N(Q(\nu),N,\alpha_s) &=&X(N,\alpha_s)(1+Q^2(F)\frac{3\alpha_{em}(0)}{4\pi})
\nonumber\\
X(N,\alpha_s)&=&  1+ \frac{C(N)}{C(3)}\frac{\alpha_s}{\pi}
\end{eqnarray}

\noindent Here $Q(F)$ is the electromagnetic charge of final state fermion
and
 $N$ refers to color representation.  The values of Casimir operators have
been
 listed earlier.  The problematic feature is the nonrationality of the
variable  $\alpha_s/\pi$. This results from the sum over the momenta of
final state   gluons on 'real  side'.  Actually these gluons are not
monitored and sum must be done 'on p-adic side'
 so that rational number results.  Instead of $\pi$   a rational number
near $\pi$ and expressible as a finite sum of powers of 2   is assumed to
result from
 summation.

\vm

\noindent d)
 $Z^0$  decay rate is affected  by the presence of
unmonitored exotic leptons as well as exotic U type quarks.
 For leptons  the value of $\alpha_s (L,M_Z)$ can be evaluated
by requiring that coupling constant diverges, say,  at $2^{-19}M_Z$. This
gives $\alpha_s (L,M_Z) \sim 0.0013$.  For quarks one can use value near
the LEP  value $\alpha_s (M_Z) = 0.115 \pm .005$.  The
 p-adic sum over amplitudes squared over exotic leptons  depends also
on    the
value of effective Weinberg angle, whose LEP value is $0.2234$.

\vm

\noindent   e) The total decay rates to exotic channels  for single
fermion generation can be derived from the knowledge of colored
excitations of quarks and leptons.  Assuming that top quark and its exotic
partners do not contribute to the exotic decays  one has for  single
fermion generation

\begin{eqnarray}
\sum \vert M(W\rightarrow L^{10}\bar{\nu}_L^{10}+10\leftrightarrow
\bar{10})\vert ^2
 &=&
20 X(10,\alpha_s(L))Y(W)p^2\nonumber\\
 \sum \vert M(Z\rightarrow exotic  \ neutrinos)\vert ^2 &=&
(54N(0,27) +20N(0,10,\alpha_s (L)) ) Y(Z)p^2 \nonumber\\
\sum \vert M(Z\rightarrow exotic  \ charged \ leptons)\vert ^2 &=&
20(1-4P_{eff}+8P_{eff}^2)N(1,10,\alpha_s(L))  Y(Z)p^2  \nonumber\\
\sum \vert M(Z\rightarrow exotic  \ quarks )\vert ^2 &=&
\sum_{N=15,15',6,24}
N N(\frac{2}{3},N,\alpha_s) (1-\frac{8P_{eff}}{3}) Y(Z)p^2 \nonumber\\
\end{eqnarray}

\noindent
 f) The    total decay width for $W$ boson is not a measured  directly
but  deduced from the total width
 of $Z^0$.
 so that the   only single experimental constraint comes  from the total
 $Z^0$ decay width.
   $Z^0$   decay amplitudes to exotic  leptons and quarks squared should
separately sum to
a very small contribution.  Assuming for  Weinberg angle to
 the  LEP value $P=0.2324\pm 0.0005$  and by choosing leptonic color
coupling
strength to be  $\alpha_s (L,M_Z)= 0.0010518$  one has

\begin{eqnarray}
\sum \vert M(W\rightarrow \  exotics )\vert^2 &=& 4.202p^2\equiv
 0.202p^2\nonumber\\
\sum \vert  M(Z\rightarrow exotic \ leptons)\vert^2&=& 45.00002p^2
\equiv 0.00002p^2
\nonumber\\
\
\end{eqnarray}

\noindent The  value of $\alpha_s (L)$ is in accordance with
the estimate obtained from evolution equation with three families
of exotic bosons.   $W$ total  unmonitored width corresponds
approximately to single $L\bar{\nu}_L$ channel and this prediction
provides  a test of the  scenario.
Also the decay width  to the $U\bar{U}$ type  exotic  pairs should be small
if these states are light. Miraculously,
for $\alpha_s(m_Z) = .116035$, which is within experimental uncertainties
 equal
 to
LEP value  $\alpha_s (m_Z)= 0.115\pm .005$,
  the decay width is of
order $10^{-5}$  if final state contains  1,2 or 3 generations of exotic
 $U$ quarks. Frustratingly, this miracle is not needed if colored quark
and boson exotics suffer primary condensation  at level $M_{89}$!

\vm

Exotic leptons poses  should be  seen in $e^+e^-$ annihilation, too. At
low energies the coupling via one photon intermediate state vanishes due
to the conservation of  vector current but at  relativistic energies the
situation is different.  The production cross amplitude squared  for
$F\bar{F}$ pair  in QED  at
 relativistic energies reads as

\begin{eqnarray}
\vert M(e^+e^-\rightarrow F\bar{F}\vert ^2 &=& \frac{e^4}{16}
(1+cos^2(\theta))
\end{eqnarray}

\noindent where $\theta$ is the angle between incoming and outgoing
fermion.  The real  counter part of the p-adicized  expression yields
enormous  production
 rate since p-adic rational number is in question.
 The correct expression must contain additional power of $p$ and does it
by previous definition of S-matrix

\begin{eqnarray}
\vert M(e^+e^-\rightarrow F\bar{F},p-adic)\vert ^2 &=&
 \frac{e^4}{16}(1+cos^2(\theta))p\simeq .52\cdot 10^{-3}(1+cos^2(\theta)) p
\nonumber\\
\
\end{eqnarray}

\noindent   If $e^2$ is finite sum of powers of $2$ (same
 applies to $cos^2(\theta)$ )  the real counterpart of this expression is
essentially the real production
 amplitude
 squared.

\vm

 At  first glance it  seems that  modulo effects cannot make production
rate small since the total p-adic  rate  to  exotic  unmonitored final
states taking into account all exotics  except top
 is only by a factor  $110  $ larger than the rate to single exotic
channel.  On the other
 hand,  p-adic unitarity requires that total p-adic  transition rate must
vanish.  The crucial question is the following: does the p-adic total
transition rate vanish
 separately for exotic final states and ordinary final states? If so then
exotic channels  sum up to zero.
 The large value of the color correction factor in $e^+e^-$  annihilation
rate is in  first order approximation  given by

\begin{eqnarray}
\vert M\vert^2 &\rightarrow &X(N)\vert M\vert^2 \nonumber\\
 X(N)&=&1+
\frac{C(N)}{C(3)}\frac{\alpha_s(L)}{\pi}
\end{eqnarray}

\noindent  provides an attractive mechanism for realizing p-adic unitary
for exotic channels separately in the
 energy region,  where $\alpha_s (L)$ becomes large.   For ordinary
$e^+e^-$ final
 states  scattering   the vanishing of sum should  due to the p-adic
summation over
 momentum degrees of freedom and the large value of the transition
amplitude squared
 in forward direction coming from Coulomb scattering should  play central
role in the mechanism.

\vm

To summarize,  the  p-adicized decay amplitudes squared  for intermediate
gauge bosons are of order  $O(p^2)$. This means that total p-adic decay
rate vanishes, when  individual decay channels
 are not monitored.  The decay rates to single channel are so near to
their maximum  values that for a p-adic  sum over several nonmonitored
channels  modulo effects become important.   The p-adic  sum of $Z^0$
decay rates   to  unmonitored exotic colored
 channels is sensitive to radiative corrections  and   with a fine tuned
value  of effective Weinberg angle it  vanishes.

\subsection{Colored excitations of leptons }

One can imagine  several scenarios explaining the anomalous $e^+e^-$ pairs
if
 one gives up the  assumption that consituent fermions for leptohadrons
are massless states and
 allows tachyonic constituents.   If only massless states are allowend
then there is only one
  scenario, which assumes light exotic elementary fermions and requires
p-adic probability theory  to
 cope with  the constraints coming from intermediate gauge boson decay
widths.  The table shows that
 leptons allow decuplet color  representations as massless states.
Neutrinos allow also two $27$ dimensional massless representations and   U
type quarks allow decuplet representations.  Therefore one might think
that color excited leptons  correspond to  $10,\bar{10},2\times 27$ rather
than octet representations according to  the original belief.

\vl

\begin{tabular} {||c|c|c|c|c||}\hline \hline
fermion & D(0)        & D(1)& D(2) &
$\frac{M_R}{m_e \sqrt{\frac{M_{127}}{p}}}$ \\
\hline  $e^{10}$ &2  & 12  & 40 & $\sqrt{\frac{9}{5+\frac{2}{3}}}$  \\
\hline
 $\nu^{10},\nu^{\bar{10}}$ & 12& 40 & 80& $  1  $ \\ \hline
$\nu^{27}$ & 2& 12 & 40&   $\sqrt{\frac{9}{5+\frac{2}{3}}}$    \\ \hline
 $U^{10},U^{\bar{10}}$& 2& 12&40& $\sqrt{\frac{9}{5+\frac{2}{3}}}$ \\
\hline\hline \end{tabular}

 \vl

Table 2.\label{Exodege} Degeneracies and masses of colored exotic fermions.

\vl

The values of leptopion masses for decuplet representation (color magnetic
and Coulombic forces are not taken into account)  are:

\begin{eqnarray}
m(\pi_L^-)&=& \sqrt{\frac{14+\frac{2}{3}}{5+\frac{2}{3}}}
m_e\sqrt{\frac{M_{127}}{p}}\nonumber\\
 m(\pi_L^0)&=&
\sqrt{\frac{18}{5+\frac{2}{3}}}  m_e\sqrt{\frac{M_{127}}{p}}\nonumber\\
\
\end{eqnarray}

\noindent The mass of the leptohadron assumed to explain  anomalous
$e^+e^-$ production should be roughly equal to  $2m_e$:   pseudoscalar
nature of the resonance allows the original identification as leptopion
 or as the leptonic counterpart of $\eta$ meson.   The precise prediction
of
 leptopion mass is not possible at this stage but one can develop
consistency arguments.

\vm

\noindent a)  Hadronic case suggests that topological mixing makes
$s(\nu^{10})$ and  $s(e^{10})$ identical and therefore not smaller than
$s_{min}=s(e^{10})=9$.   The simplest assumption is  $s_{eff}(\nu^{10})=9$
as in hadronic case: this implies that  the mixing matrix for $e^{10}$ is
trivial and for $\nu^{10}$  the matrix could be equal  to the
 matrix $D$.  for D-type quarks since the change in mass squared is same.
The absolute upper bound for for leptopion mass comes as

\begin{eqnarray}
m(\pi_L)&<&\sqrt{ \frac{2s_{eff}(e^{10})}{5+\frac{2}{3}}  }  m_e\ge
 \sqrt{\frac{36}{17} }m_e
<2m_e\nonumber\\
\
\end{eqnarray}

\noindent  One could   argue that some mixing must take place also in
charged sector and in minimal scenario one would have
$s_{min}(\nu^{10})=10$.

 \vm

\noindent
  b) If  $s(\pi_L)$ vanishes as it does for ordinary pion (the p-adic
counterpart of PCAC is that    pion  is massless  in order
 $O(p)$). With this assumption    the mass of leptopion is
 $  m(\pi_L)= \sqrt{2/17}m_e\simeq 0.343 m_e$, which  is by a factor of
order $3$ smaller than the mass of the observed resonance. In this
scenario  the decay of $e^-\rightarrow \pi^-_L+\nu_e$  via the emission of
virtual $W$ boson becomes possible and world would consist of  protons of
charged leptopions! Therefore leptopion mass must be larger than $m_e$ and
this implies $s(\pi_L)>1$. The minimal value $s(\pi)_L=6$ implies
$m(\pi_L)\simeq \sqrt{(6+2/3)/(5+2/3)}m_e$.  This is achieved if color
binding and spin spin interaction energy contribution satisfies the
 $\vert \Delta s \vert >18-5=13$ ($\Delta s=-18$ for pion).

\vm

 \noindent c)  Spin spin interaction energy is
 expected to be much smaller than in hadronic case since it is inversely
proportional to the third power of average quark-quark  distance. If
spin-spin interaction energy vanishes in first order only color Coulombic
contribution
 remains. This contribution is inversely proportional to average
quark-quark distance and simple scaling is expected to hold true and
$\Delta s_c=-9$  should hold true  for leptohadrons, too.   With this
assumption one obtains

\begin{eqnarray}
m(\pi_L)&\simeq &   \sqrt{\frac{2s_{eff}-9 +\frac{2}{3}}
{5+\frac{2}{3}}}m_e
 \end{eqnarray}

\noindent The observed resonance can be identified as  leptopion  only
provided one has $ s_{eff}(e^{10})\ge 15$, which implies  that $e^{10}$
suffers at least same amount of mixing as $u$ quark.

\vm

 \noindent d)   The alternative identifaction  of resonance as
lepto-$\eta$  is possible if  leptopion has minimal mass ($s(\pi_L)=6$ and
color magnetic  spin-spin interaction is of
 second order.  $\eta$  mass resuls from mixing only and   if the mixing
of $\eta_L$
 is identical with the mixing of $\eta$ then one has $s(\eta_L)=
s(\eta)+s(\pi_L)=6+s(\pi_L)\ge 20$, which gives $s(\pi_L)\ge 14$ and
requires
 that topological mixing is large $s_{eff}(e^{10})\ge 12$.
     In hadronic case $\eta$  however mixes with  and $\eta_c$.
 Even a very  small mixing with the leptonic counterpart of $\eta_c$ could
change  the mass of $\eta_L$ considerably  since the only possible
candidate  for the condensation level of lepto-c is $k=113$. Generalizing
from the hadronic case
 the contribution of $\eta_c^L= c_L\bar{c}_L$ to $s_{eff}(\eta)$  is of
order

\begin{eqnarray}
 \Delta s&=& p\cdot  s(\eta^c_L)\nonumber\\
s(\eta_c^L) &\simeq & 2s(c_L)12^{14} = 24 \cdot 2^{14}\nonumber\\
s(c_L)&=&12 \ \  (no \ topological \ mixing)
\end{eqnarray}

\noindent  Correct order of magnitude corresponds to $\Delta s\leq 11$
and this is achieved with extremely small  mixing fraction  $p\simeq
2^{-15}$.

\subsection{Experimental signatures of leptohadrons}

The couplings of leptomeson to electroweak gauge bosons can be  estimatd
using PCAC and CVC hypothesis \cite{Itzykson}.   The effective
$m_{\pi_L}-W$ vertex is  the matrix element of electroweak axial current
between vacuum and charged leptomeson state
 and can be deduced using same arguments as in the case of ordinary
charged pion

\begin{eqnarray}
 \langle 0 \vert J^{\alpha}_A \vert \pi_L^-\rangle= K m(\pi_l)p^{\alpha}
\nonumber\\
\end{eqnarray}

\noindent  where $K$ is some numerical factor and $p^{\alpha}$ denotes the
momentum of leptopion.
 For neutral leptopion the same argument gives vanishing coupling to photon
by the conservation of vector current. This has the important consequence
that  leptopion cannot be observed  as  resonance in $e^+e^-$
annihilation  in single photon channel.  In two photon channel leptopion
should appear as resonance.  The
 effective interaction Lagrangian is the 'instanton' density of
electromagnetic field \cite{Heavy,Lepto}  giving additional contribution
to the divergence of the
 axial current and was used to derive a model for leptopion production  in
heavy ion collisions.

\subsubsection{ Leptohadrons and lepton decays}

  The  lifetime of the  resonances decaying to $e^+e^-$ pair (neutral
leptopion)  is  known to
 be of  order $10^{-10}$ seconds:  the estimate of lifetime
\cite{Lepto,Heavy} using generalization of PCAC hypothesis is of same
order of magnitude. A rough  estimate for the  lifetime of charged
leptopion is obtained by scaling  the life time of the  ordinary  charged
pion, which is proportional to $m(\pi)^3$: this gives value
  $\tau (\pi_L^-) \sim 10^{-2}\ seconds$. If leptopion mass is very near
to electron mass its decay
 is phase space supressed and  factor increasing the life time appears. In
any case  leptopions are practically stable particles and can appear in
the final states of particle reactions.  In particular, leptopion atoms
are possible and  by Bose statistics have the peculiar property that
ground state can contain many leptopions.

\vm

 Lepton decays $L \rightarrow \nu_{\mu}+ H_L$, $L=e,\mu,\tau$  via
emission of virtual $W$ are kinematically  allowed and an anomalous
resonance  peak in the neutrino energy  spectrum at  energy

\begin{eqnarray}
E(\nu_{L})&=&\frac{m(L)}{2} -\frac{m_H^2}{2m(L)}
\end{eqnarray}

\noindent provides a  unique test for the leptohadron hypothesis.  If
 leptopion is too light   electrons  would decay to charged leptopions and
neutrinos unless the condition $m(\pi_L)>m_e$  holds true.

\vm

   The existence of  a  new  decay channel for muon  is an obvious danger
to the leptohadron scenario:
 large changes in muon decay rate are not allowed. \\
  a)   Consider first the decay  $\mu\rightarrow \nu_{\mu}+\pi_L$ where
$\pi_L$ is  on mass  shell leptopion.
  Leptopion has energy $\sim m(\mu)/2$ in muon rest system  and is highly
relativistic so that in  the muon rest system the
 lifetime  of leptopion is
 of order $\frac{m(\mu)}{2m(\pi_L)}\tau(\pi_L) $  and the average length
traveled
 by leptopion before decay is of order $10^8$ meters! This means that
leptopion
 can be treated as stable particle.  The presence of a new decay channel
changes the lifetime of muon although the  rate for events using $e\nu_e$
pair as  signature is not changed.  The effective  $H_L-W$ vertex was
deduced above.
 The rate for the decay via leptopion emission and its ratio to ordinary
rate
 for muon decay are given by

\begin{eqnarray}
\Gamma (\mu \rightarrow \nu_{\mu} + H_L)&=& \frac{G^2K^2}{2^5
\pi}m^4(\mu)m^2(H_L)(1-\frac{m^2(H_L)}{m^2(\mu)})
\frac{(m^2(\mu)-m^2(H_L))}{(m^2(\mu)+m^2(H_L))   }\nonumber\\
\frac{\Gamma (\mu \rightarrow \nu_{\mu} + H_L)}{\Gamma (\mu \rightarrow
 \nu_{\mu}+e+\bar{\nu}_e)} &=& 6 \cdot (2\pi^4) K^2
\frac{m^2(H_L)}{m^2(\mu)}\frac{(m^2(\mu)-m^2(H_L))}{(m^2(\mu)+m^2(H_L))   }
\nonumber\\
\
\end{eqnarray}

  \noindent and is of order $.93  K^2$ in case of leptopion.  As far as
the determination of $G_F$  or equivalently $m_{W}^2$ from muon decay rate
is considered the situation seems to be good since  the change
 introduced to $G_F$ is of order $\Delta G_F/G_F \simeq  0.93 K^2$ so that
$K$ must be considerably smaller than one.

\vm

  Leptohadrons can appear also as virtual particles  in the decay
amplitude $\mu\rightarrow \nu_{\mu}+e\nu_e$ and this changes the value of
muon decay rate. The correction is however extremely small since the decay
vertex of  intermediate off mass shell leptopion is proportional to its
decay rate.

\subsubsection{ Leptopions and beta decay}

 If leptopions are allowed as final state particles leptopion emission
provides a new   channel  $n\rightarrow p+ \pi_L$ for  beta  decay of
nuclei since the invariant mass of virtual  $W$ boson varies within the
 range $(m_e=0.511 \ MeV,m_n-m_p=1.293 MeV$.  The resonance peak for
$m(\pi_L) \simeq 1 \ MeV$ is extremely sharp due to the long lifetime of
the charged leptopion. The energy of  the leptopion at resonance is

\begin{eqnarray}
E(\pi_L)&=& (m_n-m_p)\frac{(m_n+m_p)}{2m_n}+\frac{m(\pi_L)^2}{2m_n}
\simeq m_n-m_p
\end{eqnarray}

\noindent Together with long lifetime this
 leptopions escape the detector volume without decaying (the exact
knowledge of the energy
 of charged leptopion  might make possible  its direct detection).

\vm

 The contribution of leptopion to neutron decay rate is not negligible.
Decay  amplitude is proportional to
 superposition of divergences of axial and vector currents between proton
and neutron states.

\begin{eqnarray}
M&=&  \frac{G}{\sqrt{2}} Km(\pi_L) ( q^{\alpha}V_{\alpha}+
 q^{\alpha}A_{\alpha})
\end{eqnarray}

\noindent    For  exactly conserved vector current the contribution of
vector current vanishes identically. The matrix element of the  divergence
of axial vector current at small momentum transfer (approximately zero)
can be evaluated using PCAC hypothesis.

\begin{eqnarray}
q^{\alpha}A_{\alpha}&=& f(\pi) m^2(\pi) \pi
\end{eqnarray}

\noindent relating the divergence to pion field,  which gives for the
 matrix element between nucleon states at zero momentum transfer  the
estimate

\begin{eqnarray}
\langle p \vert q^{\alpha}A_{\alpha}  \vert n\rangle&=& -f(\pi)
\bar{u}_p\gamma_5u_n\nonumber\\
\end{eqnarray}

\noindent  Since leptopion mass $q^2$ is very small as compared to pion
mass
  this gives estimate for divergence of the axial vector current in the
amplitude
  $\pi_L$ production rate in neutron decay.

\vm

 Straightforward calculation shows that the ratio for the decay rate  via
leptopion emission and ordinary beta decay rate is given by

\begin{eqnarray}
\frac{\Gamma(n\rightarrow
p+e+\bar{\nu}_e)}  {\Gamma(n\rightarrow p+\pi_L)}&=&
\frac{15K^2\pi^2}{2^3 cos^2(\theta_c) (1+3\alpha^2)0.47  }
\frac{m(\pi_L)^2f(\pi)^2m_n}{\Delta^5} \sqrt{1-
\frac{m^2(\pi_L)}{\Delta^2 }}\nonumber\\
&\simeq & 1.5336 \cdot 10^7K^2\nonumber\\
\Delta&=& m(n)-m(p)\nonumber\\
f(\pi)&\simeq& 93 \ MeV
\end{eqnarray}

\noindent where the parameter $\alpha$ \cite{Okun} appears in the matrix
element of
charged electroweak current

\begin{eqnarray}
J_{\mu}&=& \bar{u}_p\gamma_{\mu}(1+\alpha \gamma_5)u_n\nonumber\\
\alpha &\simeq& 1.253
\end{eqnarray}

\noindent Leptopion contribution is smaller than ordinary contribution
if the condition

\begin{eqnarray}
K&<&  10^{-4}
\end{eqnarray}

\noindent  is satisfied.  This condition implies that the relative
contribution to muon decay rate is below $10^{-8}$ and therefore
negligibly small.  Clearly,
 leptopion emission can give quite large contribution to neutron  decay
rate.

   \section{Scaled up copies of hadron Physics?}

TGD suggests the existence of scaled up copies of hadron physics
corresponding to the Mersenne  primes $M_{n}=89,61,31,..$.  The
requirement of unitarity forces the existence of Higgs in gauge theories
and since there seems to be no room for
 Higgs
 in TGD  these copies of hadron physics might guarantee unitarity. As
already
 found the increase of color coupling strength  implied by the existence
of color excited quarks
 means the end of 'old' QCD and a probable emergence of new QCD.

 \subsection{The observed top candidate and $M_{89}$
Physics?}

 The TGD:eish prediction for the top mass  for $k=89$
 and $k=97$
levels and are
$m_t(89)\simeq 871 \ GeV $ and  $ m_t(97) \simeq 54.4 \ GeV $ to be compared
with the mass  $ m_t(obs)\simeq 174 \ GeV$ of the observed top candidate.
 The study of CKM matrix in previous paper  led to the cautious conclusion
that only  the mass of the experimental  top candidate is consistent with CP
breaking  observed in
 $K-\bar{K}$  and $B-\bar{B}$ system.  A  possible resolution of discrepancy
is
the mixing of condensate levels: observed top  corresponds to the actual top,
for which small mixing of condensate  level  $k=97$ with condensate level
$k=89$ takes place:

\begin{eqnarray}
t&=& cos(\Phi)t_{97}+sin(\Phi)t_{89}\nonumber\\
sin^2(\Phi) &=&\frac{ m_t^2-m_t^2(97)}{ m_t^2(89)-m_t^2(97)}\sim .036
\end{eqnarray}

\noindent The value of the mixing angle is rather small and means that top
spends less than 4 per cent of its time on $k=89$ level.   The nice feature
of the explanation is that all primes $k=107,103,97$ below $k=89$ define
primary condensation levels for quarks.
 A much less probable  possibility is that the
observed top belongs to $M_{89}$ physics and there is top quark to be found
with mass $54.4 \ GeV$ or $871 \ GeV$.

\vm

The anomalies reported in the production rate and decay characteristics of
the top candidate might have explanation in terms of $M_{89}$ hadron
production \cite{Abe,Abachi} taking place besides the production of
$\bar{t}t$ pairs: the more recent data \cite{Abe1,Abachi1} show still
slightly
too high production cross section.
 The point is that $u$ and $d$ quarks of
$M_{89}$ hadron physics should   have masses quite near to the mass of top
candidate.  The failure to distinguish between actual top and exotic quarks
would lead to anomalous features in production and decay characteristics.
     $M_{89}$ hadrons hadron physics is
obtained in good approximation  by scaling the ordinary hadron physics by
the ratio $\sqrt{\frac{M_{89}}{M_{107}}}$.  This  implies  QCD $\Lambda$,
string tension, etc.    get scaled by the appropriate power of this
factor.
 If $g=0$ quark ($u$ or $d$ is in question) one might expect that  this
quark is observed in the decay of  scaled up $\rho_{89}$ or $\omega (89)$
meson (the mass ratio of $\omega$ and $ \rho$ is $1.02$ so that the
prediction
for the effective $u(89)$  mass does not change appreciabely).
 If one defines top mass as $m(u_{89})= m(\rho_{89})/2$ one obtains the
 TGD:eish prediction for its mass as $m(u_{89})= 512 \rho_{107}\simeq 197 \
GeV$,  which is about  $11$ per cent larger than the mass of the top
candidate. Defining $u(89)$  mass by  scaling the mass of ordinary $u$ quark
defined as one third of proton mass one obtains $u_{89}$ mass about
$8$ per cent too smaller than the mass of top candidate.

\vl

\begin{tabular}{||c|c||c|c||}
 \hline\hline
meson&$ m/GeV$&baryon &$ m/GeV$\\ \cline{1-4}\hline\hline
$\pi^{0}$ &69.1&$p$ &480.4\\ \hline
$\pi^+$ &71.5&$n$ &481.0\\ \hline
$K^+$ &252.8&$\Lambda$ &571.2\\ \hline
$K^0$ &254.8&$\Sigma^+$ &609.0\\ \hline
$\eta$ &281.0&$\Sigma^0$ &610.4\\ \hline
$\eta^,$& 490.5&$\Sigma^-$ &610.5\\ \hline
$\rho$& 394.2&$\Xi^{0}$ &673.2 \\ \hline
$\omega$& 400.9&$\Xi^-$ &676.5\\ \hline
$K^*$& 456.7 &$\Omega^-$ &856.2\\ \hline
$\Phi$&522&&\\ \hline \hline
\end{tabular}

\vl

Table 3.\label{M89masses}  Masses of low lying hadrons for $M_{89}$ hadron
 physics obtained by scaling ordinary hadron masses by a factor of $512$.

\vm

There is strong mathematical and physical   motivation for $M_{89}$ hadron
 physics. Higgs particle  is absent from TGD:eish Higgs mechanism.
Originally Higgs particle  was introduced to obtain unitary amplitudes in
massive gauge theory. It seems that some particles propagating in loops
are needed to achieve unitarity in TGD framework, too.  It is not
difficult to guess    that $M_{89}$  hadrons are the TGD:eish  counterpart
of Higgs.

\subsection{What the New Physics could look like?}

$M_{89}$ physics means the emergence of a new condensate level
 in the hadronic physics.  One can visualize  $M_{89}$ hadrons as very tiny
objects  possibly condensed on  the quarks and gluons of $M_{107}$ hadron
physics.   The New Physics begins to reveal itself, when the collision
energy
is so high that  $M_{89}$ hadrons inside quarks and gluons can exist as
on mass
shell particles ($M_{89}$ hadron inside $ M_{107}$ hadron is   comparable
to a bee  of size of one cm in a room of size about $5$ meters!).
   The new Physics at the energies not much above the energy scale of top
is essentially the counterpart of ordinary hadron physics at cm energies
of the order of $\rho/\omega$  meson mass.  Therefore $M_{89}$  meson
 resonances and their interactions described rather satisfactorily by the
old fashioned  string model with string tension scaled by factor $2^{18}$
should describe the situation. The electroweak interactions should be in
turn describable using generalization of current algebra ideas, such as
PCAC and vector dominance model.   If $M_{89}$ hadrons condense on quarks and
gluons  this physics must be convoluted with the distribution functions of
$M_{89}$ hadrons inside quarks and gluons.   The resonance structures are
partially smeared out by the convolution process.

\vm

Although the original identification of top quark candidate as $u_{89}$ and
$d_{89}$ is probably not correct it is worth of describing the proposed
scenario.
 The observation that the top candidate decays to b quark via W
emission
 has  explanation in this picture although at first this decay seems
difficult to understand since topology change  $g=0\rightarrow g=2$ seems
to be involved.  \\
a)  An essential assumption is that the outer boundary of p-adic primary
condensation level corresponds is the carrier of quark quantum numbers.
This assumption is not in accordance with the original idea that boundary
components carrying elementary particle quantum numbers have size of order
Planck length and implies that family replication phenomenon is associated
with the outer boundary of primary p-adic condensate level. One cannot
however  exclude this possibility since conformal
(and scaling) invariance is basic property
of boundary component dynamics. \\
b) Besides this one must assume that $k=89$ quarks are condensed on ordinary
quarks and gluons to explain the decays of top candidate to b-quark.
If  the $\rho$ ($I=1$) or $\omega$ ($I=0$) meson of
the
 New Physics is condensed on $b$ quark or $c$ quark then  the
disappearence of $\rho$ leaves  just $b$ or $c$ quark independently of the
reaction mechanism.\\    c) If there is no preference for condensation
level then the decay of $M_{89}$ hadrons takes with same probability via
any quark.   If the  condensation takes   place hierarchically then
$M_{89}$   quarks    of  the New Physics  prefer to condense  on $k=103$
surface  rather than to $k=107$ then decays to   $b$  quark or $c$ quark
dominate.   The scenario is  obviously excluded if top is found to decay
mostly
to b-quark.

\vm

A much more simpler picture is that boundary components carrying quark
quantum numbers are much smaller than the size of primary condenate level
forces the identification of observed top candidate as actual top and
$M_{89}$ Physics can only explain the anomalies in production and decay
of top.

\vm

 Consider next the estimation of the  production and decay rates
for
 $\rho (89)$ /$\omega (89)$ and more generally $M_{89}$ mesons.  The basic
idea is that  different condensation levels communicate only via
electroweak interactions  was already suggested in \cite{padTGD}, where
various consequences of Mersenne hierarchy of Physics  were considered.
The basic interaction is the emission of electroweak gauge boson followed
by the decay of the gauge boson to   $M_{89}$ quark pair condensed on
ordinary quark or gluon  and the reverse of this process.   Electroweak
gauge boson  $B$ can be produced  either in $q\bar{q}$ annihilation to
gluon plus boson:
 $q+\bar{q}\rightarrow B+ g$ or in Compton
 scattering $ q+g\rightarrow q+B$.  Compton scattering dominates in the
energies considered.  The ratio of $\alpha_{em}/\alpha_{s}\simeq.046$
implies that about the cross section is about $5$ per cent from that for
the production of ordinary top quark.

 \vm

Since low energies are in question at $M_{89}$ level the scaled up version
of
 vector dominance model  described in the nice  book of Feynmann
\cite{Feynmann} should give a  satisfactory description for the production
of $M_{89}$ mesons via resonance mechanism. The idea is to introduce
direct coupling $F_V= m_V^2/g_V$  of photon (or gauge boson)  to vector
boson  ($\rho$, $\omega$, $\phi$). The diagrams describing the production
of mesons via decay of vector boson contain vector boson propagator
$\frac{1}{p^2-m_V^2+ im_V\Delta}$ and   the production rate is enhanced by
a factor $ R=4\pi m_V^2/(\Delta^2g_V^2)$ in the resonance: the factor
should be same in $M_{89}$ physics as in ordinary hadron physics.  The
ratio $r=\alpha_{em}R/\alpha_s$ gives a rough measure for the ratio of the
rates of production for $u(89)$ and ordinary top quark.  A rough estimate
for what is to be expected is obtained by scaling the results of ordinary
hadron physics. The table below gives the estimates for the quantity $r$
and one has $r=15.1$ for $\omega$.   $\omega$ is  clearly a more probable
candidate for the resonance structure  observed in Fermilab.

\vl

\begin{tabular}{|| c|c|c|c|c||}\hline\hline
meson& $m/512 \ MeV$ &$ \Delta/512\ MeV$ &$g_V^2/4\pi$ &$r$\\
 \cline{1-5}\hline
$\rho$ &770&  150&2.27&0.52\\ \hline
$\omega$ &783&10&18.3&15.1\\ \hline
$\Phi$&1019&4.2&13.3&230.8\\ \hline\hline
\end{tabular}

\vl

Table 4. \label{Resprodparam} Scaled up resonance production parameters
for  $\rho$, $\omega$ and $\Phi$. The last column of the table gives the
value of the  quantity $r=\alpha_{em}R/\alpha_s$, which should give a
measure   for  the ratio of  production rate of $u(_89)$  and of  the the
production of ordinary top quark pair.

\vm

 The proton of $M_{89}$ physics cannot decay to ordinary quarks if only
 electroweak interactions are allowed. $M_{89}$ proton is stable unless
there
 is a mechanism for the transfer of fermion number between different
levels of topological condensate!   A possible TGD:eish mechanism for this
transition  is topological evaporation \cite{padTGD}.  The quarks of
$M_{89}$ hadron evaporate coherently (color singletness) from $M_{89}$
condensate level and condenses back to $M_{107}$ level.  The understanding
of $B$ and $D$ meson masses is also based on the possibility of
topological evaporation.  In \cite{padTGD} a model for topological
evaporation and condensation based on p-adic length scale hypothesis and
dimensional arguments was developed but contains several unnecessary ad hoc
assumptions.

\vm

 An alternative mechanism is based on
the idea that different p-adic topologies correspond to different phases and
phase transitions changing the value of $p$ are possible. In present case
the phase transition changing $k=89$ topology to $k=97$ topology starts from
small seed of $k=107$ topology, which grows and fills the entire $k=89$
hadron volume. Similar phase transition takes place on the primary
condensation level of quarks,  too. The decay of $M_{89}$ hadrons is
described as proceeding via the decay to virtual $M_{89}$ hadrons regarded
as
super cooled $k=89$ phase. Virtual $M_{89}$ mesons can decay electroweakly
and
$k=89$ baryons suffer eventually phase transition to $k=107$ phase and
resulting ordinary quark gluon plasma decays to ordinary hadrons.

\vm

One of the  suggested applications of
topological evaporation \cite{padTGD} was the explanation of Centauro type
events
 \cite{Centauro} in terms slightly different rates for the  coherent
topological evaporation of quarks and antiquarks, which makes possible the
situation, when quarks are in vapour phase and antiquarks in condensate or
vice versa.
  A much simpler
description free of ad hoc assumptions seems  however to be possible. The
point
is that  virtual $M_{89}$ mesons  can decay electroweakly and $M_{89}$ pions
produce photon pairs with energies differing widely from the energies of the
ordinary pions: thus the anomalously small abundance of ordinary pions. The
decay of virtual $M_{89}$ baryons via phase transition to ordinary baryons
and mesons.

\vm

Topological evaporation provides  an explanation for the  mysterious
 concept of
 Pomeron  originally introduced to describe hadronic diffractive
scattering  as the exchange of  Pomeron Regge trajectory \cite{Pomeron}.
No hadrons belonging to  Pomeron
 trajectory were however found and via the advent of QCD Pomeron was
almost  forgotten.  Pomeron  has
 recently experienced reincarnation \cite{Hera,pantip,pp}.  In Hera
\cite{Hera}     $e-p$    collisions,   where proton scatters essentially
elastically whereas jets in the direction of incoming
 virtual photon emitted by electron  are observed.    These events can be
understood  by assuming   that proton emits
 color singlet particle  carrying small fraction of proton's momentum.
This particle  in turn  collides with virtual photon (antiproton)   whereas
proton scatters essentially elastically.
 The identification of the color singlet particle as pomeron looks natural
since pomeron emission describes nicely  diffractive scattering of
hadrons.   Analogous hard diffractive scattering events in  $pX$
diffractive scattering  with $X=\bar{p}$  \cite{pantip} or
 $X=p$ \cite{pp} have also been observed.  What happens is that  proton
scatters essentially elastically and emitted  pomeron collides with $X$
and suffers hard scattering so that large rapidity gap jets  in the
direction  of $X$ are observed.   These results suggest that Pomeron is
real and consists of ordinary partons.

\vm

 The TGD:eish identification of Pomeron is very economical:  Pomeron
corresponds to sea partons, when valence quarks are in vapour phase.
 In TGD inspired phemenology  events involving Pomeron correspond to $ pX$
collisions, where  incoming $X$   collides with proton, when valence
quarks have suffered coherent  simultaneous (by color confinement)
evaporation into vapour phase.  System $X$ sees only the sea left behind
in evaporation and scatters from it whereas valence quarks continue
without noticing $X$   and condense later to form quasielastically
scattered proton.   If $X$ suffers hard scattering from the  sea the
peculiar hard diffractive scattering  events are observed.  The fraction
of these events is  equal to the fraction  $f$ of time spent by valence
quarks  in vapour phase.   In \cite{padTGD} dimensional
 arguments were used to derive a rough order of magnitude estimate for
$f\sim 1/\alpha =1/137\sim 10^{-2}$ for $f$: $f$ is of same order of
magnitude as
 the fraction (about 5 per cent) of peculiar events from all deep inelastic
scattering events in Hera.    The time spent in condensate  is by
dimensional arguments of the order of the p-adic legth scale
$L(M_{107})$,  not far from  proton Compton length.   Time dilation
effects at high collision energies guarantee that valence quarks indeed
stay in vapour phase during the collision. The identification of Pomeron
as  sea   explains also why Pomeron Regge trajectory does not correspond
to actual on mass shell particles.

\vm

  The existing detailed knowledge about the properties of sea structure
functions
 provides a stringent test for the TGD:eish scenario.   According to
\cite{pantip}  Pomeron structure
 function seems to consist of soft  ($(1-x)^5$ ),   hard ($(1-x)$ ) and
superhard component  (delta function like component at $x=1$).  The
peculiar super hard component finds explanation in TGD:eish picture.  The
structure function $q_P(x,z)$ of parton in Pomeron  contains  the
longitudinal momentum  fraction $z$  of the Pomeron as a parameter and
$q_P(x,z)$ is obtained by scaling from the sea  structure function $q(x)$
for proton  $q_P(x,z)= q(zx)$.   The
 value of structure function at $x=1$ is nonvanishing: $q_P(x=1,z)= q(z)$
and this explains the necessity to introduce super hard  delta function
component in the fit of \cite{pantip}.

\vm

To sum up, $u_{89}$ and $d_{89}$ quarks have masses near to top quark mass
and
 production  and decay of  $M_{89}$ hadrons  might explain the reported
anomalies in top production and decay.    Following list gives some of the
unique signatures of New Physics.\\ a)  At higher energies exotic  pions are
produced abundantly and might be  detectable via annihilation to monoenergetic
photon  pair.   $\pi^0$ of the New Physics should have mass  $69.1 \ GeV$
and
$\gamma \gamma$ annihilation width $512\cdot 7.63 \ eV= 3.9 \ MeV$ (obtained
by scaling from that for ordinary pion).  The width for the decay by $W$
emission from either quark of $\pi^0 (89)$ (the second is assumed to act as
spectator)  is of order $ G_F^2m(u(89))^5/(192\pi^3)$  and of order $2.5$ MeV.
\\
 b) The scaling of mass splittings inside isopin multiplets with the scale
 factor $512$ as compared to ordinary hadron physics is a unique signature
of $M_{89}$ hadrons.  \\ c) The scaled up versions of $\rho$ and $\omega$
meson should be  found at nearby energies.  Kaon  (and $s$  quark) of the
New Physics should be
 seen  as a decay product of $\Phi (522 \ GeV) \rightarrow K+\bar{K}$: from
table 5. one finds that  that $\Phi$ should have rather small hadronic
width $\Delta \simeq 2.2 \ GeV$  so that the parameter measuring its
production rate to the production rate of ordinary quark is as high as
$r\simeq 230.8$ at resonance. \\ d) Since $\omega_{89}$ is superposition
of form $u_{89}\bar{u}_{89}-d_{89}\bar{d}_{89}$ half of its decays are
 of type  $d_{89}\rightarrow q+ Z^0$ rather than $u_{89}\rightarrow q+ W$.
There are  indeed indications that the decays of the  top candidate contain
anomalously too large fraction of $Z^0$ decays whereas $W$ decays are
supressed.

\vl

\begin{center}
{\bf Acknowledgements\/}
\end{center}

\vm

It would not been possible to carry out this work without the  concrete
help  of
 my friends  in concrete problems of the everyday life and I want to
express my gratitude to  them.  Also I want to thank J. Arponen,
 R. Kinnunen, J.
 Maalampi, J. Ignatius and P. Ker\"anen
    for practical help and interesting discussions.

\end{document}